\documentclass[preprint,floatfix]{revtex4}
\usepackage{graphicx}
\usepackage{epsfig}
\usepackage{enumerate}
\usepackage{amssymb}
\usepackage{amsmath}
\usepackage{amsfonts}
\usepackage{dcolumn}
\usepackage{bm}
\usepackage{color}
\usepackage{siunitx}
\usepackage{layout}
\usepackage{soul}

\begin{document}
\begin{center}

\textbf{\large{Quasi-phase-matched up- and down-conversion in periodically poled layered semiconductors}}
\vspace{0.5cm}

Chiara Trovatello$^{1,2,*}$, Carino Ferrante$^3$, Birui Yang$^4$, Josip Bajo$^{5,6}$, Benjamin Braun$^{5,6}$, Zhi Hao Peng$^1$, Xinyi Xu$^1$, Philipp K. Jenke$^{5,6}$, Andrew Ye$^7$, Milan Delor$^8$, D. N. Basov$^4$, Jiwoong Park$^{7,9,10}$, Philip Walther$^{5,11}$, Cory R. Dean$^{4}$, Lee A. Rozema$^{5,*}$, Andrea Marini$^{3,12}$, Giulio Cerullo$^{2,13,*}$ and P. James Schuck$^{1,*}$

\vspace{0.5cm}

\textit{\footnotesize{
$^1$Department of Mechanical Engineering, Columbia University, New York, NY 10027, USA\\
$^2$Dipartimento di Fisica, Politecnico di Milano, Piazza L. da Vinci 32, I-20133 Milano, Italy\\
$^3$CNR-SPIN, c/o Dip.to di Scienze Fisiche e Chimiche, Via Vetoio, Coppito (L’Aquila) 67100, Italy\\
$^4$Department of Physics, Columbia University, New York, NY 10027, USA\\
$^5$Faculty of Physics, Vienna Center for Quantum Science and Technology (VCQ), University of Vienna, Boltzmanngasse 5, 1090 Vienna, Austria\\
$^6$Faculty of Physics \& Vienna Doctoral School in Physics, University of Vienna, Boltzmanngasse 5, 1090 Vienna, Austria\\
$^7$Pritzker School of Molecular Engineering, University of Chicago, Chicago, IL, USA\\
$^8$Department of Chemistry, Columbia University, New York, NY 10027, USA\\
$^9$James Franck Institute, University of Chicago, Chicago, IL, USA\\
$^{10}$Department of Chemistry, University of Chicago, Chicago, IL, USA\\
$^{11}$Research Network Quantum Aspects of Space Time (TURIS) \& Christian Doppler Laboratory for Photonic Quantum Computer, University of Vienna, Boltzmanngasse 5, 1090 Vienna, Austria\\
$^{12}$Department of Physical and Chemical Sciences, University of L’Aquila, Via Vetoio, 67100 L’Aquila, Italy\\
$^{13}$Istituto di Fotonica e Nanotecnologie, Consiglio Nazionale delle Ricerche, Piazza L. da Vinci 32, 20133 Milano, Italy\\
}}

\vspace{0.5cm}
 
{\footnotesize \textit{$^*$Corresponding authors. Email:} chiara.trovatello@polimi.it, lee.rozema@univie.ac.at, giulio.cerullo@polimi.it, p.j.schuck@columbia.edu}

\end{center}

%%%%%%%%%%%%%%%
\newpage

\noindent \textbf{{\footnotesize{Nonlinear optics lies at the heart of classical and quantum light generation. The invention of periodic poling revolutionized nonlinear optics and its commercial applications by enabling robust quasi-phase-matching in crystals such as lithium niobate. However,  reaching useful frequency conversion efficiencies requires macroscopic dimensions, limiting further technology development and integration. Here we realize a periodically poled van der Waals semiconductor (3R-MoS$_2$). Due to its large nonlinearity, we achieve macroscopic frequency conversion efficiency of 0.03\% at the relevant telecom wavelength over a microscopic thickness of 3.4~$\SI{}{\micro m}$ (\textit{i.e.}, 3 poling periods), $10-100\times$ thinner than current systems with similar performances. Due to intrinsic cavity effects, the thickness-dependent quasi-phase-matched second harmonic signal surpasses the usual quadratic enhancement by 50\%. Further, we report the broadband generation of photon pairs at telecom wavelength via quasi-phase-matched spontaneous parametric down-conversion, showing maximum coincidence-to-accidental-ratio (CAR) of $638\pm75$. This work opens the new and unexplored field of phase-matched nonlinear optics with microscopic van der Waals crystals, unlocking applications that require simple, ultra-compact technologies such as on-chip entangled photon-pair sources for integrated quantum circuitry and sensing.\\}}}

%%%%%%%%%%%%%%%
%%%%%%%%%%%%%%%
%%%%%%%%%%%%%%%

\footnotesize

The inherent nonlinear response of matter to external electromagnetic stimuli allows photons of different frequencies incident on a material to interact with each other, enabling a myriad of photonic applications such as frequency conversion and the generation of non-classical states of light. Particularly interesting are second-order nonlinear processes, which occur in non-centrosymmetric media with $\chi^{(2)}\neq0$ and are used to produce new light frequencies \textit{e.g.}, in second harmonic generation (SHG), sum and difference frequency generation, and entangled photon pairs in spontaneous parametric down-conversion (SPDC)\cite{Boyd2020,Shih2016}.

Efficient nonlinear frequency conversion is achieved by fulfilling the so-called phase matching (PM) condition, which implies a zero wave-vector mismatch - \textit{i.e.}, momentum conservation - for the interacting waves ($\Delta k=0$). In the following we consider the case for SHG, but identical considerations apply for SPDC. For SHG, $\Delta k=k_{2\omega}-2k_\omega=\frac{2\omega}{c}(n_{2\omega}-n_\omega)$ and PM requires matching the refractive indexes of the fundamental wavelength (FW), $n_\omega$, and of the second harmonic (SH), $n_{2\omega}$. In the absence of PM, the conversion efficiency reaches its maximum for a propagation distance corresponding to the coherence length $L_c$ = $\pi$/$\Delta$k and then oscillates with period $2L_c$\cite{Boyd2020}. Birefringent phase-matching (BPM) exploits the optical anisotropy of non-centrosymmetric nonlinear crystals by adjusting the propagation direction inside the crystal  so that the fields at $\omega$ and $2\omega$ with different polarizations experience the same refractive index. 
For SHG with BPM, the SH intensity $I_{2\omega}$ grows quadratically with the thickness $L$ of the nonlinear medium. While BPM is simple and effective, it can only be applied to a limited number of crystals, such as the prototypical $\beta$-barium borate (BBO), which display moderate $\chi^{(2)}$ values of a few pm/V\cite{Boyd2020}.

There are many crystals with high nonlinearity but low anisotropy, \textit{e.g.}, gallium arsenide (GaAs), for which BPM is not achievable. An alternative to BPM is quasi-phase-matching (QPM), which introduces periodic phase shifts of $\pi$ between the fields at the FW and the SH at every coherence length $L_c$, restoring the proper phase relationship and the quadratic growth of the SH intensity with propagation length\cite{Fejer1992,Myers1995}.

Such a phase shift can be obtained by periodic inversion of the crystallographic orientation during material growth via molecular beam epitaxy\cite{Koh1998,Eyres2001,Grisard2012}, or by cleaving the nonlinear medium slabs along the different crystal planes and diffusion-bonding them into monolithic stacks\cite{Gordon1993,Grisard2012,Tanimoto2021}.
In ferroelectric crystals like lithium niobate (LN)\cite{Feng1980,Feisst1985}, lithium tantalate \cite{Matsumoto1991} or potassium titanyl phosphate \cite{VanDerPoel1990}, QPM can be achieved by periodic poling, \textit{i.e.}, by periodically inverting the sign of the nonlinear coefficient $\chi^{(2)}$ upon high-voltage switching of ferroelectric domains following lithographic patterning of the electrodes\cite{Hum2007}. The spatial modulation of $\chi^{(2)}$, which follows the electrode spacing, is the so-called poling period, and it determines the frequencies for which a certain nonlinear process can be phase matched. For LN, the typical values of the poling periods are between 5 and $\SI{50}{\micro m}$ and the modulation can be extended over crystal thicknesses from millimeters to centimeters\cite{Hum2007}.

The invention of QPM was a breakthrough in nonlinear optics because it enabled the use of nonlinear crystals with large $\chi^{(2)}$ of the order of $30-\SI{50}{pm/V}$, for which BPM cannot be achieved\cite{Boes2023}. Periodically poled LN (PPLN) crystals provide the highest conversion efficiency for waveguided SHG ($I_{2\omega}/I_\omega\approx70\%$)\cite{wang2018,suntsov2021watt}, optical parametric amplifiers and oscillators\cite{Myers1997,Yang1999,lu2021,Ledezma2023} and entangled photon sources\cite{Solntsev2018,Zhang2021}. Along with their high conversion efficiencies, standard nonlinear crystals like PPLN can be directly fabricated on silicon-based optical circuits, though their millimeter-to-centimeter thickness limits the number of  devices that can be integrated on a single chip. This has spurred recent advances in the development of nonlinear metasurfaces\cite{Krasnok2018,Wang2022,Fedotova2022,Santiago-Cruz2022,Neshev2023}, thin films\cite{Zhu2021} and nanowaveguides\cite{Jankowski2023}, which are now leading a paradigm shift towards more miniaturized, and more tunable, nonlinear optical device platforms.

The rise of quantum and two-dimensional (2D) material science\cite{Guo2023,Wu2017,Mueller2018,Du2023,Ma2023,Sheffer2023} has inspired research on less traditional classes of optical crystals including semiconducting transition metal dichalcogenides (TMDs), \textit{e.g.}, MoS$_2$, MoSe$_2$, WS$_2$ and WSe$_2$. These are layered materials comprised of crystalline sheets with strong in-plane covalent bonds but weak out-of-plane van der Waals interactions. Starting from the first reports of SHG\cite{Malard2013,Li2013,Wang2015shg}, their exceptional optical nonlinearities ($\chi^{(2)}=100-1000~$pm/V), up to $100\times$ higher than standard bulk crystals due in part to favourable quantum geometries, have recently been exploited to demonstrate nonlinear effects at the ultimate thin-ness limit in single-layer TMDs\cite{Autere2018,Wen2019,Liu2020,Dogadov2022}. However, in such cases, the overall conversion efficiency is still limited by the sub-nm propagation length\cite{trovatello2021optical}. In contrast to the most common 2H crystallographic phase, which is centrosymmetric in the bulk ($\chi^{(2)}=0$), the 3R polytype is non-centrosymmetric ($\chi^{(2)}\neq0$) and can be used to boost the conversion efficiency of second-order nonlinear optical processes by increasing the number of layers $N$ in the TMD sample. Using thin 3R-MoS$_2$ flakes (1-6 layers), it has been shown that the SH intensity grows quadratically with the number of layers, \textit{i.e.}, with the thickness of the nonlinear medium ($I_\textrm{SHG}\propto N^2$)\cite{Zhao2016}.

However, when the number of layers is increased to $\sim200-300$, the wave vector mismatch $\Delta k$ results in an inevitable deviation from the $N^2$ enhancement\cite{Shi2017,Xu2022}, and the SH oscillates with the sample thickness, with a semi-period equal to the coherence length $L_c$\cite{Boyd2020}. Recently, the coherence length $L_c$ in 3R-MoS$_2$ has been measured at telecom wavelengths\cite{Xu2022}. At $L_c$ ($\sim\SI{500}{nm}$) the SHG efficiency is $1.5 \times 10^4$ higher than for a monolayer ($I_{2\omega}/I_\omega\approx10^{-6}$), resulting in the highest conversion efficiency per unit length reported to date for a transparent material\cite{Xu2022}. Further increasing this value to achieve macroscopic nonlinear conversion efficiencies, while still preserving the micrometer thickness of TMD  crystals, remains an open challenge.

Here we bridge this gap using a non-centrosymmetric van der Waals crystal, 3R-MoS$_2$, to achieve QPM in periodically-poled transition metal dichalcogenides (PPTMDs) at the relevant telecommunications wavelengths. We flip the sign of the optical nonlinearity $\chi^{(2)}$ by stacking consecutive slabs with opposite dipole orientation, obtaining programmable microstacks with thickness-tunable phase-matching bandwidths. The large nonlinearity of TMDs combined with QPM unlocks single-pass conversion efficiencies for SHG of 0.03\% at telecom wavelengths over only 3 poling periods, \textit{i.e.}, a microscopic thickness of $\SI{3.4}{\micro m}$. We observe cavity effects prompted by internal reflections of both FW and SH, which increase the conversion efficiency by an additional 50\% over standard QPM. Finally, we report efficient broadband generation of photon pairs at telecom wavelengths via quasi-phase-matched SPDC, showcasing a maximum coincidence-to-accidental-ratio (CAR) of $638 \pm 75$ which outperforms by almost two orders of magnitude any microscopic SPDC source based on van der Waals crystals reported to date\cite{Guo2023,Weissflog2023ArXiV}.\\

%%%%%%%%%%%%%%%
%%%%%%%%%%%%%%%
%%%%%%%%%%%%%%%

\begin{figure}[h!]
\includegraphics[width=1\textwidth]{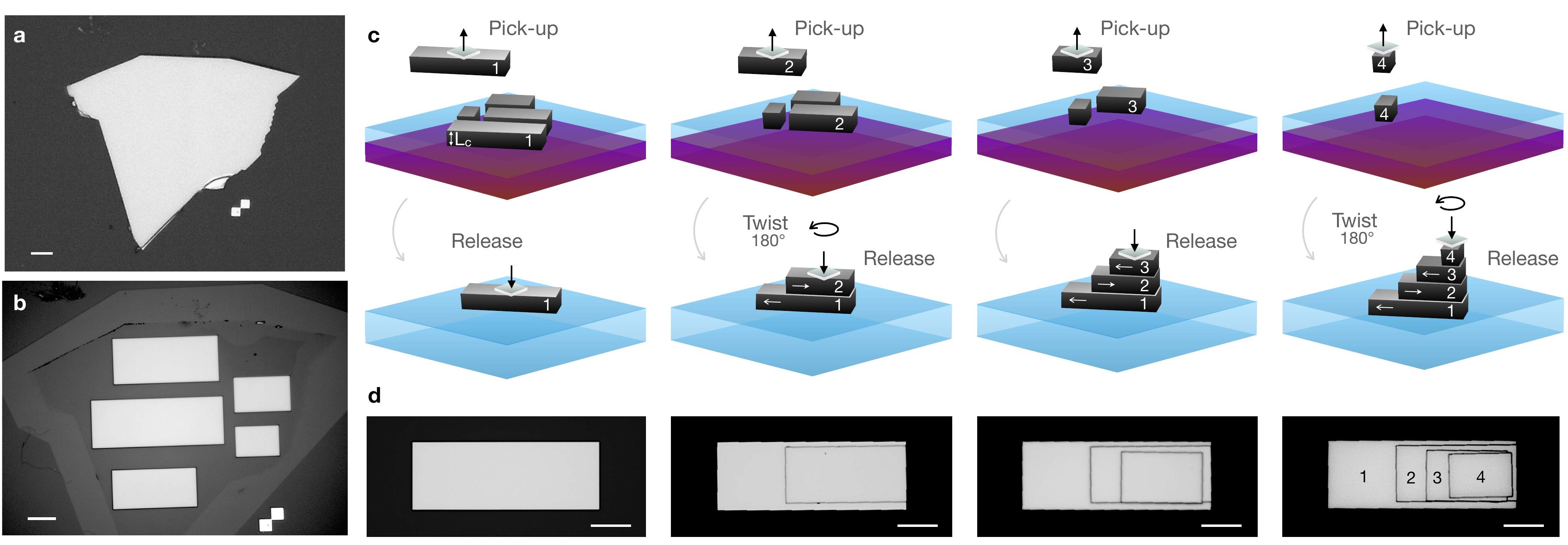}
\centering
\caption{\scriptsize{\textbf{Periodically poled transition metal dichalcogenides (PPTMDs).} \textbf{a-b}, Micrograph of the 3R-MoS$_2$ flake (\textbf{a}) before and (\textbf{b}) after patterning (electron beam lithography, etching). \textbf{c}, Stacking procedure. Slab 1 is first transferred on a $\SI{500}{\micro m}$-thick SiO$_2$ transparent substrate. Slab 2 is twisted by $\SI{180}{\degree}$ and released on top of slab 1. Slab 3 is transferred on top of slab 2. Finally slab 4 is twisted by $\SI{180}{\degree}$ and released on top of the 3 stacked portions. \textbf{d}, Micrographs of each stacking step. Scale bar $\SI{10}{\micro m}$.}}
\label{fig:sampleprep}
\end{figure}

\noindent The 3R-MoS$_2$ crystals are mechanically exfoliated from a commercial bulk 3R-MoS$_2$ crystal (HQ graphene) grown by chemical vapor transport on a SiO$_2$/Si substrate. Characterization of the bulk  crystal by energy dispersive X-ray analysis and X-Ray diffraction is provided in Ref.\cite{Xu2022}. The thickness of the exfoliated flakes is measured using atomic force microscopy (Supplementary Note 4). To fabricate the PPTMD, we select a large flake with lateral size of $\approx\SI{100}{\micro m}$ and thickness $\approx\SI{300}{nm}$, shown in Fig. \ref{fig:sampleprep}a. We pattern the flake into smaller portions (Fig. \ref{fig:sampleprep}b, Supplementary Note 1 for details) using a modified CMOS process, \textit{i.e.}, electron beam lithography followed by reactive ion etching (RIE), the standard in large scale device production. TMDs are readily dry etched with SF$_6$-based RIE, which does not require hard-mask and achieves high-quality side walls (see AFM profiles of the patterned slabs in Supplementary Note 4). By cutting the different slabs out of a single flake, we ensure that all areas have identical thickness and the same macroscopic dipole orientation. Our design and fabrication methodology thus bypasses the need for angle-resolved SHG for crystal orientation characterization.

We choose a flake with thickness of $L_{\rm f}\approx\SI{300}{nm}$, close to the coherence length $L_{\rm c}$ measured for a FW at 1450 nm (Supplementary Note 7). The largest portion of the flake (area 1) is transferred onto a $\SI{500}{\micro m}$-thick transparent fused silica (SiO$_2$) substrate, and the other portions (area 2, 3, 4) are individually stacked on top of each other by keeping an interlayer twist angle of $180^\circ$\cite{ArxivTwistPM2023} (symmetrically equivalent to $60^\circ$ and $300^\circ$ twist angles) to flip the sign of $\chi^{(2)}$ at each coherence length (see Fig. \ref{fig:sampleprep}c), resulting in an overall thickness of $\approx\SI{1.2}{\micro m}$ (Supplementary Note 4). The microscope images of the periodically poled crystal at each stacking step are shown in Fig. \ref{fig:sampleprep}d.

\begin{figure}[h!]
\includegraphics[width=1\textwidth]{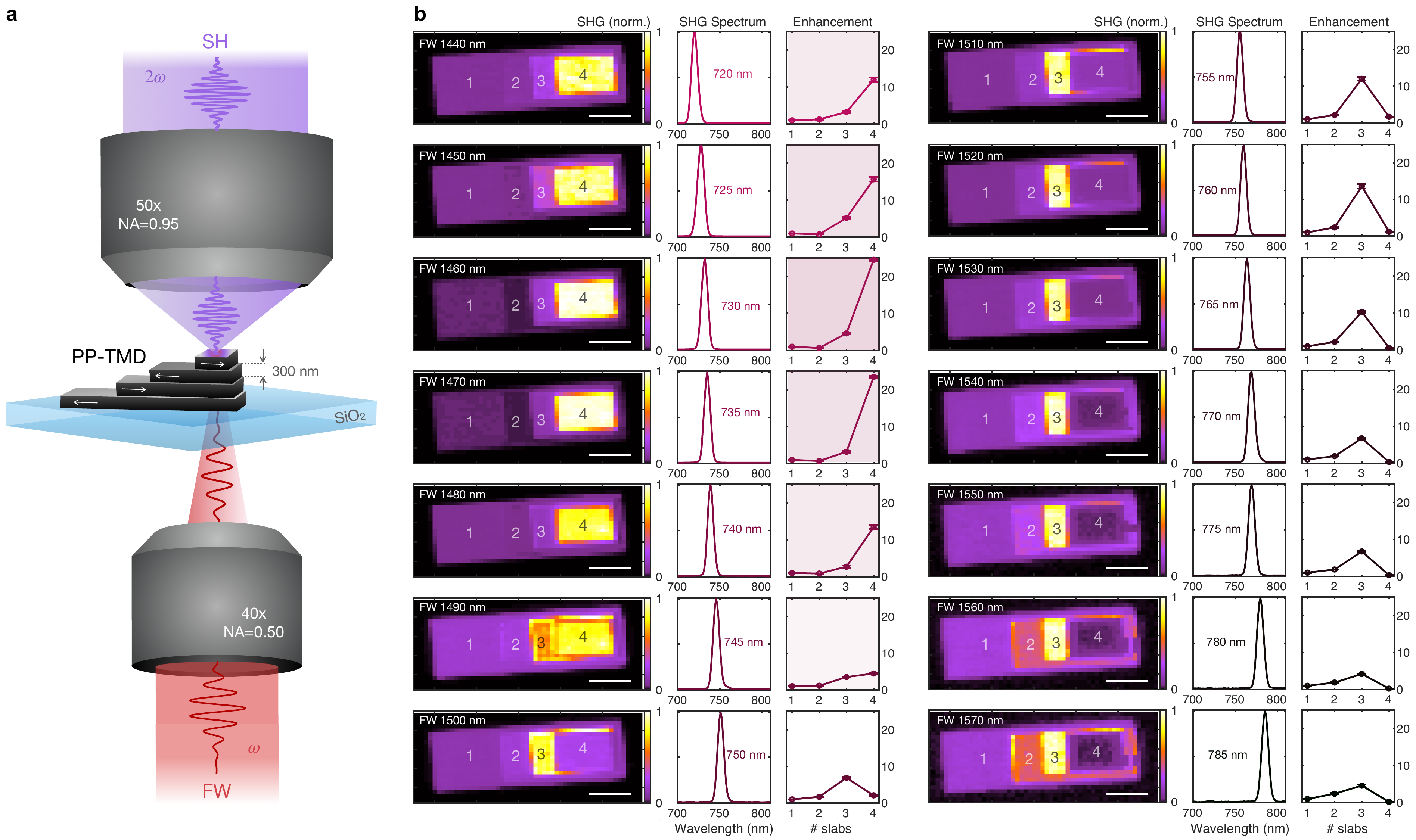}
\centering
\caption{\scriptsize{\textbf{Quasi-phase-matched  second-harmonic generation (SHG) from PPTMDs.} \textbf{a}, Transmission microscope. The PPTMD is excited by the FW from the back side of the SiO$_2$ substrate with a $40\times$ objective. The SH is collected by a $50\times$ objective. \textbf{b}, Pump wavelength-dependent normalized SH maps of the PPTMD, along with the SH spectra and the SHG enhancement factor with respect to the bare slab. The error bar represents the standard deviation of the SH intensity over 20 pixels (1 pixel = $\SI{1}{\micro m}\times\SI{1}{\micro m}$) of each corresponding slab area. Data are presented as mean values of the SH intensity over 20 pixels of each corresponding slab area. Scale bar $\SI{10}{\micro m}$.}}
\label{fig:wl_dep}
\end{figure}

To characterize the nonlinear response of the PPTMD, we use a custom-built transmission microscope (Fig. \ref{fig:wl_dep}a), illuminated by an optical parametric oscillator (OPO, Coherent Chameleon), emitting $\SI{200}{fs}$ pulses at $\SI{80}{MHz}$, tunable in the NIR (1000-1600 nm) wavelength range. We use a long working-distance $40\times$ reflective objective with numerical aperture (NA) 0.5 to focus the FW on the sample from the backside of the substrate. The SH is collected by a $50\times$ objective with a 0.95 NA and directed onto a silicon CCD camera (Supplementary Note 2). The FW is tuned from 1430 nm to 1580 nm, and the power is kept fixed at $\SI{0.5}{mW}$ (peak power $\SI{1.17}{GW/cm^2}$). Figure \ref{fig:wl_dep}b shows the pump wavelength dependent normalized SHG intensity maps, along with the corresponding SHG spectrum.
For each pump wavelength, we extract the SH enhancement factor, defined as the SH emission intensity from the regions with 2, 3 and 4 slabs normalized to the SH emission intensity of 1 slab, shown in Fig. \ref{fig:wl_dep}b. The peak of the SH enhancement is obtained at FW $=\SI{1460}{nm}$, at approximately the target operation wavelength for this stack. Compared to a standard first-order QPM ($\chi^{(2)}$ flipped in sign at each coherence length) that predicts a quadratic enhancement of 4, 9 and 16 for 2, 3 and 4 slabs, respectively, the peak enhancement that we obtain is not monotonically increasing with the slab number at each pump wavelength, and it is almost 50\% higher in the portion of the stack with 4 slabs. This effect can be explained by considering the modulation of the optical properties caused by interference inside the PPTMD\cite{Shoji97}. Owing to the large refractive index of 3R-MoS$_2$, ranging between 4 and 5.5\cite{Xu2022}, multiple reflections of FW and SH radiation at the interfaces between the TMD and the air and glass substrate, acting as optical microcavities, provide nonlinear cavity enhancement effects.\\

\begin{figure}[h!]
\includegraphics[width=1\textwidth]{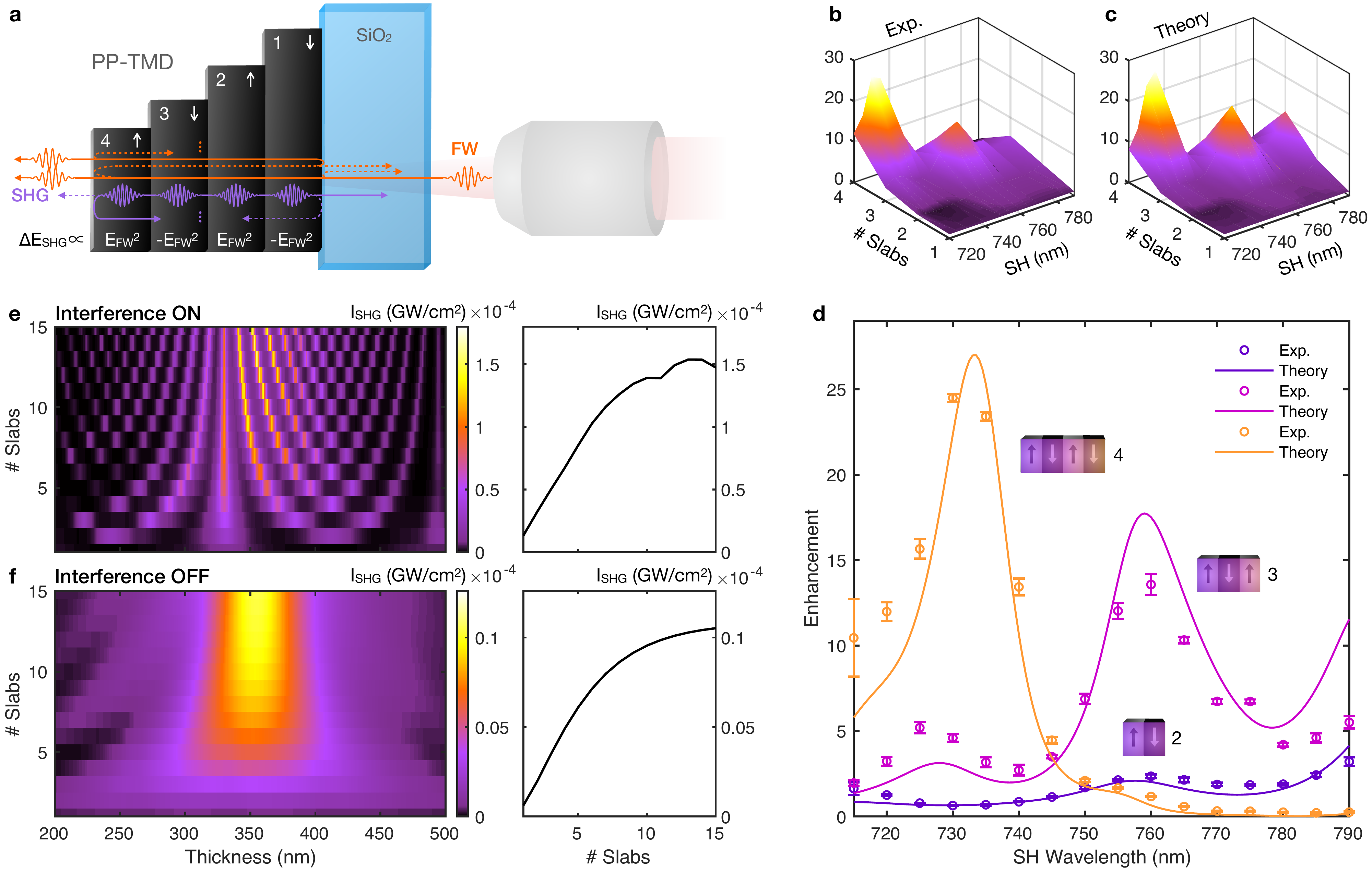}
\centering
\caption{\scriptsize{\textbf{Theoretical simulations of quasi-phase matching in PPTMDs.} \textbf{a}, Schematic of simulated PPTMD structure. Multiple reflections of FW and SH fields are shown in orange and purple, respectively. The inversion of the electric field at each coherence length is induced by $\chi^{2}$ flipped in sign via twist-controlled stacking of consecutive slabs. \textbf{b}, SH enhancement for different wavelengths and number of slabs, as reported in Fig. \ref{fig:wl_dep}. \textbf{c}, Corresponding SH enhancement extracted from the model. \textbf{d}, Quantitative comparison of the wavelength-dependent enhancement between experiment and model for different number of slabs, \textit{i.e.}, cross sections of panel (b) and (c), respectively. The error bar represents the standard deviation of the SH intensity over 20 pixels (1 pixel = $\SI{1}{\micro m}\times\SI{1}{\micro m}$) of each corresponding slab area, as extracted in Fig.~\ref{fig:wl_dep}. Data are presented as mean values of the SH intensity over 20 pixels of each corresponding slab area. \textbf{e-f}, Calculated SH intensity as a function of slab thickness and number of slabs at FW = 1450 nm and $\SI{5}{GW/cm^2}$ intensity, in two cases: with interference effects of the FW (\textbf{e}), and without interference (\textbf{f}). The maximum intensity achievable as a function of the total number of slabs in the PPTMD is shown on the right, showing that interference boosts SHG, due to intrinsic cavity-induced enhancements.}}
\label{fig:th}
\end{figure}

\noindent To better understand the unconventional QPM regime we observe in PPTMDs, we analytically model our signal by solving the coupled nonlinear equations considering the interference of FW and SH fields in the slabs, and the SH electric field sign inversion for the different poling conditions (Supplementary Note 6).
Specifically, we assume that the SH process does not affect the intensity of the FW (undepleted-pump approximation) and we apply the boundary conditions at the entrance and the exit of the TMD slabs to analytically retrieve the interference effects. 
Calculating the forward and backward propagating FW electric fields and considering the poling, we are able to evaluate the SH in the system. We calculate the second order nonlinear polarization in the slabs at $2\omega$ induced by FW propagation, and insert this term into Maxwell's equations to extract the SH electric and magnetic fields.
Applying again the boundary condition for the SH, we obtain the forward emitted SH intensity. The analytical calculation is performed assuming a normal propagation in the different slabs. A sketch of the modeled structure is shown in Fig. \ref{fig:th}a, in which the interference effects of FW and SH are depicted.

Figure \ref{fig:th}b and \ref{fig:th}c show the measurements and the theoretical simulations of the SH enhancement factor (\textit{i.e.}, SH emission from slab 2, 3 and 4 normalized to the SH from slab 1) as a function of the SH wavelength and the number of slabs, respectively. For the theoretical simulations, we use $|\chi^{(2)}|=~\SI{100}{pm/V}$ and a slab thickness equal to $\SI{293}{nm}$ (Supplementary Note 6). Figure \ref{fig:th}d reports the comparison between experiments and theory for poled structures with 2, 3 and 4 slabs, demonstrating a very good agreement with the data. The theoretical model also accurately reproduces the nonlinear response of PPTMDs with different poling periods (Supplementary Note 8). Data and simulations emphasize the importance of an appropriate choice of the number of slabs for each FW wavelength and a given slab thickness, to maximize the enhancement of SH. In particular, for our 300-nm slab thickness, a configuration with 3 slabs provides a higher enhancement compared to the poled structure with 4 slabs for SH wavelengths above 750 nm.

Figure \ref{fig:th}e shows the intensity profile of SH for different slab thicknesses and number of slabs. The results show a maximum efficiency for a $L_{\rm f}\sim\SI{350}{nm}$, with a strong dependence on the interference effects of the FW. The optimum poling period is strongly dependent on the FW. For instance, at 1550 nm the maximum efficiency is achieved for $L_{\rm f}\sim\SI{550}{nm}$ (see Supplementary Note 6).
We also report the same simulation removing the interference of the FW in the slab (Fig. \ref{fig:th}f), which we achieve by setting the real part of refractive index of air after the slabs equal to that of the TMD. In this case, the interference pattern is removed, obtaining a more homogeneous profile. However, the maximum efficiency is notably reduced (more than $10\times$ lower).
This comparison highlights the key role played by the intrinsic cavity enhancement effects in PPTMDs.\\

\begin{figure}[h!]
\includegraphics[width=.75\textwidth]{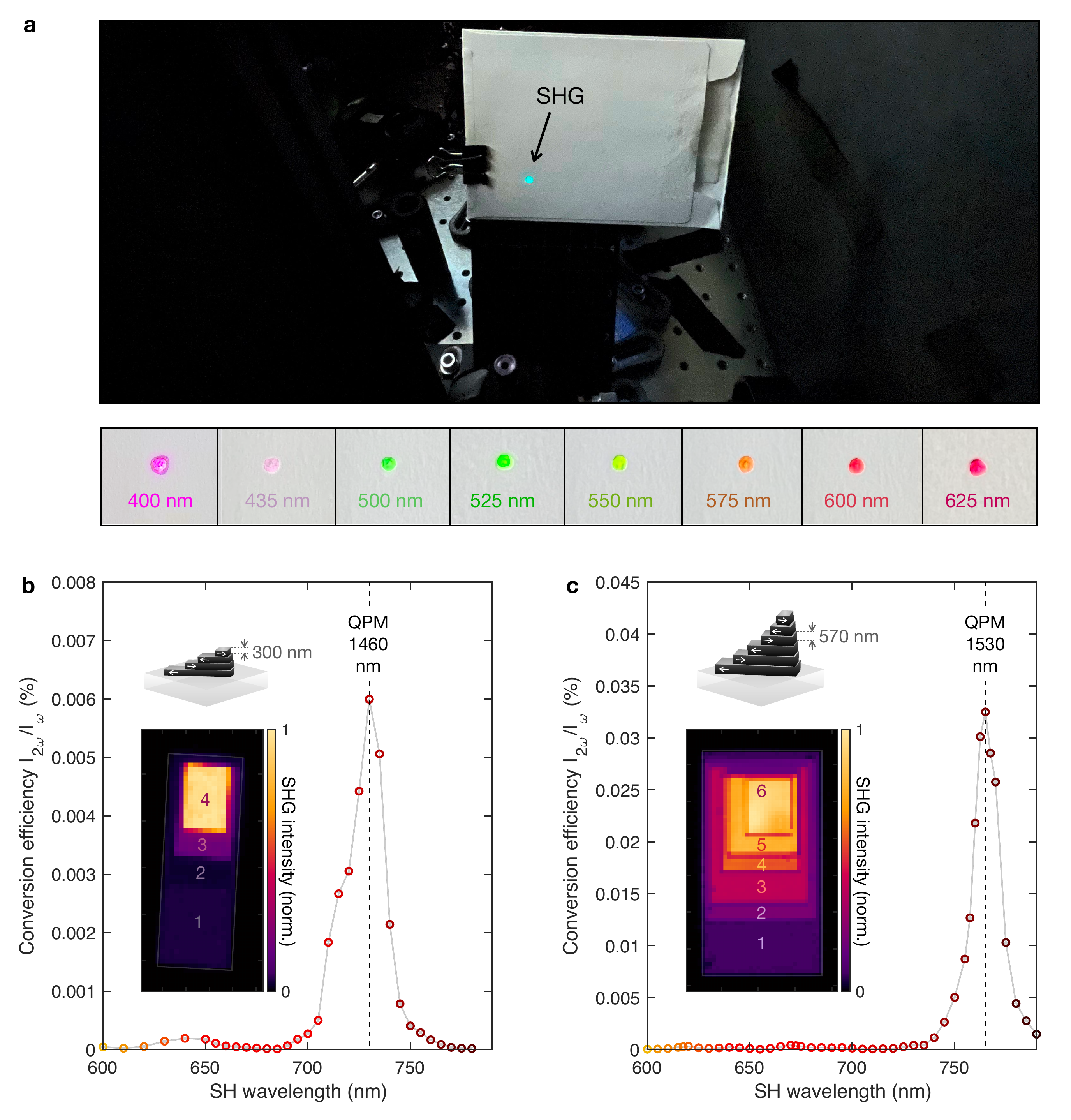}
\centering
\caption{\scriptsize{\textbf{PPTMD second harmonic conversion efficiency.} \textbf{a}, Picture of the SHG spot at 530 nm, and close-up of the tunable SH from 400 nm to 625 nm. \textbf{b-c}, Broadband SH conversion efficiency ($I_{2\omega}/I_\omega$) with tunable pump wavelength (1200 nm - 1590 nm) measured on the (b) PPTMD with slab thickness 300 nm (QPM resonance 1460 nm) and (c) PPTMD with slab thickness 570 nm (QPM resonance 1530 nm). The pump power (b) 40 mW (peak power $\SI{98}{GW/cm^2}$) (c) 52 mW (peak power $\SI{127}{GW/cm^2}$). In the insets exemplary normalized SHG intensity maps of the PPTMDs. Scale bar $\SI{10}{\micro m}$.}}
\label{fig:efficiency}
\end{figure}

\noindent To quantitatively measure the conversion efficiency of our PPTMD samples we measure the pump wavelength dependent SH power. We use the frequency tunable OPO as laser source. The photographs of the broadly tunable SH spots emitted from the $\SI{1.2}{\micro m}$-thick PPTMD are shown in Fig. \ref{fig:efficiency}a. Owing to the macroscopic efficiencies of our PPTMDs (the SH reaches $\approx\SI{10}{\micro W}$ powers at the phase-matching bandwidth, see Supplementary Note 9), the SH power has been measured using a standard, spectrally calibrated, silicon power meter (Thorlabs S120VC). To show that we can realize highly efficient and programmable microstacks with tunable phase matching, we measure the SH power emitted by two different PPTMD samples: the 4-stack PPTMD (2 poling periods) with slab thickness of 300 nm, and a second 6-stack PPTMD (3 poling periods) with slab thickness of 570 nm (micrograph of the 6-stack PPTMD and SH spatial mapping are shown in Fig. S8, Supplementary Note 9). Each sample is excited with tunable pump wavelength and a constant pump power. Figure 4b and 4c show the conversion efficiency, \textit{i.e.}, SH power / FW power, of the PPTMD with slab thicknesses $\SI{300}{nm}$ and $\SI{570}{nm}$, respectively. The QPM resonance is peaked at 1460 nm (Fig. 4b) and 1530 nm (Fig. 4c), respectively, demonstrating QPM resonance tunability with the slab thickness, as well as validating the theoretical model that predicts, with very good agreement, the optimum slab thickness as a function of the FW wavelength (see Fig. S4, Supplementary Note 6).

The peak of the conversion efficiency approaches $\approx10^{-4}$ (\textit{i.e.}, 0.01\%) at FW $=\SI{1460}{nm}$ for the 4-stack PPTMD (Fig. 4b) and 0.01-0.1\% at FW $=\SI{1530}{nm}$ for the 6-stack PPTMD (Fig. 4c). We stress the fact that such macroscopic conversion efficiencies are recorded over a sample thickness of only $\SI{1.2}{\micro m}$ and $\SI{3.4}{\micro m}$, respectively. PPTMDs eclipse the previous thickness-conversion efficiency trade-off curves, and now show macroscopic efficiencies over microscopic thicknesses. Additionally, to retrieve the QPM bandwidth as a function of the number of slabs in the PPTMD, we measure the conversion efficiency of the PPTMD phase-matched at $\SI{1530}{nm}$ from the portion of the sample with 4 and 5 slabs, and we compare it with the emission from the 6 slabs. The extracted FWHM of the QPM bandwidths are $22.4\pm1.4$, $16.6\pm1.8$ and $15.5\pm1.0$ for the PPTMD with 4, 5 and 6 slabs, respectively (see Fig. S8 in Supplementary Note 9). At the QPM resonance in the the relevant telecom bandwidth, the extracted nonlinearity of 3R-MoS$_2$ is $\chi^{(2)}\sim\SI{100}{pm/V}$ (see Supplementary Note 10). With higher $\chi^{(2)}$ compared to BBO/PPLN\cite{Boyd2020,Shoji97}, PPTMDs achieve the same efficiency, but up to  $100\times$ shorter propagation lengths.\\

\begin{figure}[h!]
\includegraphics[width=.9\textwidth]{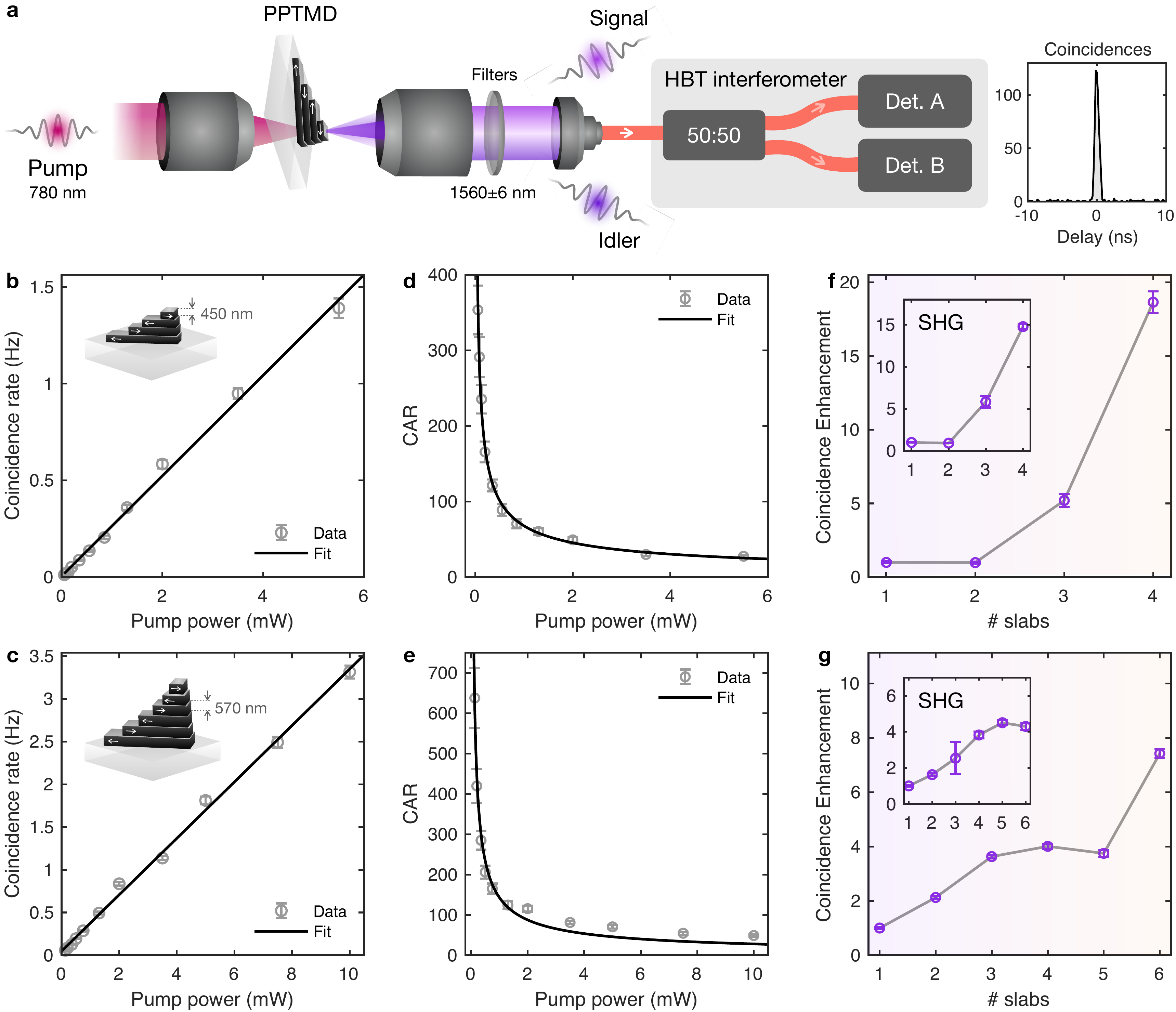}
\centering
\caption{\scriptsize{\textbf{Quasi-phase-matched Spontaneous Parametric Down-Conversion (SPDC) from PPTMDs.} \textbf{a}, Experimental setup.  A CW laser at 780 nm is used to pump the PPTMD in a transmission geometry. The generated SPDC photons are directed to a Hanbury Brown-Twiss (HBT) interferometer for photon-pair correlation measurements. On the right side, we plot the signature of photon-pair detection, \textit{i.e.}, an exemplary coincidence peak at the zero delay in the arrival time histogram between the counts of two detectors.
\textbf{b-c}, Pump-power-dependent SPDC coincidence rates (dots) for the 4-stack PPTMD (b) and 6-stack PPTMD (c), with a corresponding linear fits (black lines). Data points are presented as the coincidence rate averaged over the measurement time. The error bar is calculated as the Poissonian error of total coincidence events over the whole measurement.
\textbf{d-e}, Pump power dependence of the CAR (coincidence to accidental ratio) peak (dots) at the zero time delay for the 4-stack PPTMD (d) and 6-stack PPTMD (e), with corresponding hyperbolic fits (black lines). Data points are presented as the coincidence rate averaged over the measurement time divided by accidental coincidence events. The error bar is calculated as the Poissonian error of total coincidence events over the whole measurement divided by accidental coincidence events.
\textbf{f-g} Nonlinear enhancement of measured photon coincidence rates versus number of slabs for the 4-stack PPTMD (f) and 6-stack PPTMD (g). The pump power is set to 4 mW (f) and 12.5 mW (g). Insets in (f) and (g) show the trends in enhancement versus number of slabs of the measured SHG signal for 4-stack and 6-stack PPTMDs, respectively. Data points are presented as the coincidence rate averaged over the measurement time, normalized to the first data point. The error bar is calculated as the Poissonian error of total coincidence events over the whole measurement, normalized to the first data point (slab $\#1$).
}}
\label{fig:SPDC}
\end{figure}

\noindent Finally, we use PPTMDs to demonstrate SPDC. For these measurements, we employ two PPTMD samples quasi-phase-matched in the telecom C band. In particular, we use a 4-stack PPTMD with a slab thickness of $\SI{450}{nm}$, giving the best nonlinear enhancement at $\SI{1560}{nm}$ (see Supplementary Note 10), as well as the previously shown 6-stack PPTMD with a slab thickness of $\SI{570}{nm}$ and best enhancement at $\SI{1530}{nm}$. To perform temporal correlation measurements in the relevant telecom wavelength range we use a different experimental setup that is optimized for SPDC measurements. It features a different laser source and  different detectors in a transmission geometry, as depicted in Fig. \ref{fig:SPDC}a.

In particular, we use a continuous wave  $\SI{780}{nm}$ laser (TOPTICA) as a pump for the SPDC process. An aspheric lens (NA=0.68) and an objective (NA=0.85), optimized for telecom wavelengths, are used to focus the pump onto the sample and collect the down converted light, respectively. After the PPTMD, the pump beam is filtered out with three hard-coated long-pass filters with a cutoff wavelength of $\SI{1150}{nm}$. Additionally,  we use a hard-coated band pass filter at $\SI{1560}{nm}\pm\SI{6}{nm}$, to only collect the degenerate portion of the SPDC emission. Due to chromatic aberration in our collection optics, this enhances our coincidence-to-accidental-ratio (CAR). The SPDC signal is then coupled into a single mode fiber (SMF28, Corning). Photon pairs are probabilistically split into two different fiber paths using a fiber-based  Hanbury-Brown-Twiss interferometer (Fig. 5a), then guided to superconducting nanowire single-photon detectors (PhotonSpot), where detection events are registered by a timetagger (Universal Quantum Devices). Analyzing temporal correlations of detection events yields a timing histogram, which, when detecting photon pairs generated by SPDC, is expected to show strong correlations at zero time delay between the detectors. A  histogram taken in this experimental configuration is shown on the right of Fig. \ref{fig:SPDC}a.

Figure 5b and 5c show the coincidence rates (dots) of the 4- and 6-stack PPTMDs respectively as a function of the pump power. Due to the non-negligible background, accidental coincidence events are subtracted. All error bars are calculated as the standard deviation of the measured coincidence rates. The rates follow the expected linear dependence.

Figure 5d and 5e show the pump power-dependent CAR along with a hyperbolic fit. At the relevant telecom wavelength, we reach a maximum CAR value of $638 \pm 75$,  more than one order of magnitude higher than the previously reported values in microscopic van der Waals materials at visible wavelengths, and two orders of magnitude larger at telecom wavelengths\cite{Guo2023,Weissflog2023ArXiV}. Replacing the band-pass filters with broadband long-pass filters (with cutoffs at $\SI{1300}{nm}$ or $\SI{1500}{nm}$), coincidences still show a sharp correlation peak, indicating that the SPDC process is efficient over a broad spectral range. The corresponding data is presented in Supplementary Note 10.

In Figure 5f and 5g, the relative enhancement of the coincidence rates is plotted as a function of the number of slabs in the PPTMD samples. The coherence length of degenerate SPDC is the same as that for the analogous SHG process at the same wavelengths, \textit{i.e.}, the conversion of telecom to visible wavelengths and vice versa, which has been shown to be $\sim\SI{500}{nm}$\cite{Xu2022}. We observe that the coincidence rate  increases with the slab number (which is proportional the thickness of the medium). Since the medium thickness exceeds the coherence length, this increase can be unambiguously attributed to quasi-phase-matching. The slight deviations from a monotonic increase in efficiency can be attributed to cavity effects from the etalon-like behavior of the PPTMDs acting on the pump and the down-converted light, as also observed for SHG. The insets show the SHG enhancement converting from $\SI{1560}{nm}$ to $\SI{780}{nm}$. In the 4-stack PPTMD, with QPM resonance exactly peaked at $\SI{1560}{nm}$, SPDC and SHG enhancements are in excellent agreement (Fig. 5f). The slight difference between SHG and SPDC enhancements in the 6-stack PPTMD (Fig. 5g) may be attributed to the strong dispersion of the refractive index modulating the effective interaction length of the pump light with the material. The slab-dependent coincidence rates with broadband filtering, reported in Supplementary Note 11, also show a net increase of the coincidence rate, again indicating that the SPDC process is efficient over a broad spectral range, enabled by quasi-phase-matching.

Finally, we compare our maximum coincidence rate to the brightest bulk sources. We take, as an example, the sources that have been used to generate 10-photon entanglement\cite{wang2016experimental}.
Normalizing their reported brightness by the square of the crystal length (since the pair-generation rate scales with the interaction length squared) yields a brightness per interaction length of $\SI{3}{MHz/W/mm^2}$.
In our case, we observe a coincidence rate of $\sim 4.4$ Hz with a pump power of 0.5 mW, and a total crystal thickness of $\SI{3.42}{\micro m}$, using broadband filtering (see Supplementary Note 11).
This gives us a source brightness per length of $\sim \SI{750}{MHz/W/mm^2}$, $250\times$ larger than the best bulk SPDC sources.

Moreover, the performance of our source could be notably improved by increasing the coupling efficiency using, for example, a variety of nanofabrication technologies such as metasurfaces\cite{MetaRoadmap2024}. Although our source may be comparable or exceed bulk sources in some figures of merit, we believe that the true strength of our platform is its compatibility with integrated on-chip components\cite{Datta2020}.
In this context, producing photon pairs directly on chip will be an enabling technology for next-generation photonic quantum devices, bypassing the loss associated with coupling each photon onto the chip\cite{Bristol2019}.
This loss scales exponentially with the number of photons produced. Thus, integrating PPTMDs into this environment could address a major bottleneck in photonic quantum computing.\\

\noindent In conclusion, we introduce periodically poled van der Waals layered materials as a novel, micrometer-sized, programmable nonlinear platform for applications in classical and quantum light generation. To realize periodically poled microstacks, we develop a modified CMOS process - standard in large scale device production - which makes use of electron beam lithography and reactive ion etching to directly pattern individual 3R-MoS$_2$ flakes into smaller rectangular slabs with identical thickness equal to the coherence length. By periodically stacking the slabs with an interlayer twist angle of $180^{\circ}$, we flip the sign of $\chi^{(2)}$ and achieve quasi-phase matching, restoring the proper phase relationship between fundamental and second harmonic, and between fundamental and spontaneously down-converted fields. Although the sample has been prepared by manual exfoliation and stacking, the periodic poling process is potentially scalable, as it can be automatized using a robotic stacking machine\cite{mannixye2022} (Supplementary Note 5). 

Our programmable microstacks show tunable phase matching as a function of the different poling period, \textit{i.e.}, the slab thickness. Due to the large nonlinearity of 3R-TMDs at the relevant telecom wavelength we demonstrate record second harmonic conversion efficiencies between 0.01\% and 0.1\% over a thickness of only $\SI{3.4}{\micro m}$ (3 poling periods). Moreover, in the phase-matched interaction, we observe nonlinear enhancement which surpasses by more than 50\% the usual quadratic enhancement typically observed in standard quasi-phase-matched crystals. This is attributed to cavity effects, which enhance the field overlap inside the periodically poled structure. Theoretical simulations accurately reproduce the conditions for such an unconventional quasi-phase matching, and predict the optimal slab thickness as a function of the FW/SH wavelength (Supplementary Note 6). Additionally, we demonstrate the generation of down-converted photon pairs via SPDC at the relevant telecom wavelength, \textit{i.e.}, $\SI{1550}{nm}$. The reported coincidence-to-accidental-ratio (CAR) from PPTMDs reaches a maximum value of $638\pm75$, the highest value observed to date in microscopic van der Waals materials\cite{Guo2023,Weissflog2023ArXiV}.

We envision that the highest impact of our work will be in establishing a platform that enables on-chip technologies for new applications at the micrometer scale, \textit{e.g.}, generating entangled photons through SPDC on-chip - a goal not feasible with the existing bulk crystal systems. This microscale approach removes the losses typically involved in the coupling of SPDC photons to a chip, and the efficiencies of TMDs demonstrated here render this a promising and competitive system for on-chip quantum light generation. Additionally, being ultrathin (\textit{i.e.}, well below $\SI{1}{mm}$), PPTMDs do not introduce substantial light walk-off, and do not require sophisticated alignment procedures, making them robust and versatile platforms that are also stable in ambient conditions for months/years. Recently, other ultrathin van der Waals compounds have also shown remarkable SH efficiencies in the visibile range (FW$=\SI{800}{nm}$, SH$=\SI{400}{nm}$) over micrometer thicknesses\cite{Abdelwahab2022}, though issues with long-term stability may be present\cite{Guo2023}, further highlighting the strengths of TMDs.

We believe our programmable microstacks set the groundwork for realizing tunable phase-matched microscopic crystals with ultracompact footprints. PPTMDs now provide macroscopic nonlinear conversion efficiencies over microscopic thicknesses, establishing new routes for designing novel nonlinear optical devices and quantum nanophotonic circuit elements\cite{Elshaari2020,silva2017}, such as entangled-photon sources based on SPDC directly embedded on chip.

\section*{ACKNOWLEDGEMENTS}

\noindent We thank Benedikt Ursprung for the useful discussions.

\noindent This work was supported by Programmable Quantum Materials, an Energy Frontier Research Center funded by the US Department of Energy, Office of Science, Basic Energy Sciences, under Award DE-SC0019443.
C.T. acknowledges the European Union’s Horizon Europe research and innovation programme under the Marie Skłodowska-Curie PIONEER HORIZON-MSCA-2021-PF-GF grant agreement No 101066108. C.T. also acknowledges the Optica Foundation and Coherent Inc. for supporting this research through the Bernard J. Couillaud prize 2022. G.C acknowledges support by the Progetti di ricerca di Rilevante Interesse Nazionale (PRIN) of the Italian Ministry of Research 2022HL9PRP Overcoming the Classical limits of ultRafast spEctroSCopy with ENtangleD phOtons (CRESCENDO). C.T. and G.C. acknowledge funding from the European Union—NextGenerationEU under the National Quantum Science and Technology Institute (NQSTI) Grant No. PE00000023-q-ANTHEM-CUP H43C22000870001.
A.M. acknowledges funding from the European Union—NextGenerationEU under the Italian Ministry of University and Research (MUR) National Innovation Ecosystem Grant No. ECS00000041-VITALITY-CUP E13C22001060006, and Progetti di ricerca di Rilevante Interesse Nazionale (PRIN) of the Italian Ministry of Research PHOTO (Photonic Terahertz devices based on topological materials) No. 316 2020RPEPNH.
A.Y. acknowledges support from the Department of Defense (DoD) through the National Defense Science and Engineering Graduate (NDSEG) Fellowship Program. J.P. acknowledges funding from the Air Force Office of Scientific Research (FA9550-21-1-0323) and the Office of Naval Research (N000142212841).
P.W. acknowledges support from the Air Force Office of Scientific Research under award number FA8655-20-1-7030 (PhoQuGraph) and FA8655-23-1-7063 (TIQI). This research was funded in whole or in part by the Austrian Science Fund (FWF)[10.55776/F71]. The financial support by the Austrian Federal Ministry of Labour and Economy, the National Foundation for Research, Technology and Development and the Christian Doppler Research Association is gratefully acknowledged.
L.A.R. acknowledges support from the Erwin Schrödinger Center for Quantum Science \& Technology (ESQ Discovery).

\subsection*{AUTHOR CONTRIBUTIONS}
\noindent C.T., G.C., and P.J.S. conceived the experiment. C.T. also conceived of the design and realization of the samples and performed the measurements. C.F. and A.M. developed the theory model. J.B., B.B., C.T. and P.K.J. performed the spontaneous parametric down-conversion measurements. B.Y., C.T., Z.P., X.X. and A.Y. prepared the samples. X.X., C.T. and Z.P. built the experimental setup and performed the morphological characterization of the samples. M.D., D.N.B, J.P., L.A.R., P.W., C.D., G.C. and P.J.S. supervised the study. C.T. wrote the manuscript with input from all authors. 

\subsection*{COMPETING INTERESTS}
\noindent The authors declare no competing interests.

\section*{DATA AVAILABILITY}
\noindent All data generated or analysed during this study, which support the plots within this paper and other findings of this study, are available on the public repository Zenodo, DOI: 10.5281/zenodo.13987619.

%%%%%%%%%%%%%%%%%%
%%%%%%%%%%%%%%%%%%
%%%%%%%%%%%%%%%%%%


\begin{thebibliography}{10}
\section*{REFERENCES}

%1
\bibitem{Boyd2020} Boyd, R. W. Nonlinear Optics. \textit{Academic Press.} (2020).

\bibitem{Shih2016} Shih, Y. An Introduction to Quantum Optics
Photon and Biphoton Physics. \textit{Taylor \& Francis.} (2016).

\bibitem{Fejer1992} Fejer, M. M., Jundt, D. H., Byer, R. L. \& Magelh, G. A. Quasi-phase-matched second harmonic generation: tuning and tolerances. \textit{IEEE Journal of Quantum Electronics.} \textbf{28}, 2631 -- 2654 (1992).

\bibitem{Myers1995} Myers, L. E. et al. Quasi-phase-matched optical parametric oscillators in bulk periodically poled LiNbO3. \textit{Journal of the Optical Society of America B.} \textbf{12}, 2102--2116 (1995).

\bibitem{Koh1998} Koh, S. et al. Sublattice Reversal in GaAs/Si/GaAs (100) Heterostructures by Molecular Beam Epitaxy. \textit{Jpn. J. Appl. Phys.} \textbf{37}, L1493 (1998).

\bibitem{Eyres2001} Eyres, L. A. et al. All-epitaxial fabrication of thick, orientation-patterned GaAs films for nonlinear optical frequency conversion. \textit{Appl. Phys. Lett.} \textbf{79}, 904--906 (2001).

\bibitem{Grisard2012} Grisard, A. et al. Fabrication and applications
of orientation-patterned gallium arsenide for mid-infrared generation. \textit{Physica Status Solidi (C).} \textbf{9}, 1651-- 1654 (2012).

\bibitem{Gordon1993} Gordon, L. et al. Diffusion-bonded stacked GaAs for quasiphase-matched second-harmonic generation of a carbon dioxide laser. \textit{Electronics Letters.} \textbf{29}, (1993).

\bibitem{Tanimoto2021} Tanimoto, R., Takahashi, Y. \& Shoji, I. Quasi-phase-matching stack of 25 GaAs plates with high transmittance for high-power mid-infrared wavelength conversion fabricated by use of room-temperature bonding. \textit{Physica Status Solidi (C).} \textbf{38}, (2021).

\bibitem{Feng1980} Feng, D. et al. Enhancement of second-harmonic generation in LiNbO3 crystals with periodic laminar ferroelectric domains. \textit{Appl. Phys. Lett.} \textbf{37}, (1980).

\bibitem{Feisst1985} Feisst, A. \& Koidl, P. Current induced periodic ferroelectric domain structures in LiNbO 3 applied for efficient nonlinear optical frequency mixing. \textit{Appl. Phys. Lett.} \textbf{47}, (1985).

\bibitem{Matsumoto1991} Matsumoto, S., Lim, E. J., Hertz, H. M. \& Fejer, M. M. Quasiphase-matched second harmonic generation of blue light in electrically periodically-poled lithium tantalate waveguides. \textit{Electronics Letters.} \textbf{27}, (1991).

\bibitem{VanDerPoel1990} Van Der Poel, C. J., Bierlein, J. D., Brown, J. B. \& Colak, S. Efficient type I blue second-harmonic generation in periodically segmented KTiOPO4 waveguides. \textit{Appl. Phys. Lett.} \textbf{57}, 20 (1990).

\bibitem{Hum2007} Hum, D. S. \& Fejer, M. M. Quasi-phasematching. \textit{Comptes Rendus Physique.} \textbf{8}, (2007).

\bibitem{Boes2023} Boes, A. et al. Lithium niobate photonics: Unlocking the electromagnetic spectrum. \textit{Science.} \textbf{379}, eabj4396 (2023).

\bibitem{wang2018} Wang, C. et al. Ultrahigh-efficiency wavelength conversion in nanophotonic periodically poled lithium niobate waveguides. \textit{Optica.} \textbf{5}, 1438-1441 (2018).

\bibitem{suntsov2021watt} Suntsov, S., R{\"u}ter, C. E., Br{\"u}ske, D. \& Kip, D. Watt-level 775 nm SHG with 70\% conversion efficiency and 97\% pump depletion in annealed/reverse proton exchanged diced PPLN ridge waveguides. \textit{Opt. Express.} \textbf{29}, 11386--11393 (2021).

\bibitem{Myers1997} Myers, L. E. \& Bosenberg, W. R. Periodically poled lithium niobate and quasi-phase-matched optical parametric oscillators. \textit{IEEE Journal of Quantum Electronics.} \textbf{33}, 10 (1997).

\bibitem{Yang1999} Yang, S. T. \& Velsko, S. P. Frequency-agile kilohertz repetition-rate optical parametric oscillator based on periodically poled lithium niobate. \textit{Opt. Lett.} \textbf{24}, 133--135 (1999).

\bibitem{lu2021} Lu, J. et al. Ultralow-threshold thin-film lithium niobate optical parametric oscillator. \textit{Optica.} \textbf{8}, 539--544 (2021).

\bibitem{Ledezma2023} Ledezma, L. et al. Octave-spanning tunable infrared parametric oscillators in nanophotonics. \textit{Sci. Adv.} \textbf{9}, eadf9711 (2023).

\bibitem{Solntsev2018} Solntsev, A. S. et al. LiNbO3 waveguides for integrated SPDC spectroscopy. \textit{APL Photonics.} \textbf{3}, 10 (2018).

\bibitem{Zhang2021} Zhang, C. et al. Spontaneous Parametric Down-Conversion Sources for Multiphoton Experiments. \textit{Adv. Quantum Technol.} \textbf{4}, 2000132 (2021).

\bibitem{Krasnok2018} Krasnok, A., Tymchenko, M. \& Alù, A. Nonlinear metasurfaces: a paradigm shift in nonlinear optics. \textit{Materials Today.} \textbf{21}, 8--21 (2018).

\bibitem{Wang2022} Wang, K., Chekhova, M., \& Kivshar, Y. Metasurfaces for quantum technologies. \textit{Physics Today.} \textbf{75}, 3745--3763 (2022).

\bibitem{Fedotova2022} Fedotova, A. et al. Lithium Niobate Meta-Optics. \textit{ACS Photon.} \textbf{9}, 3745--3763 (2022).

\bibitem{Santiago-Cruz2022} Santiago-Cruz, T. et al. Resonant metasurfaces for generating complex quantum states. \textit{Science.} \textbf{377}, 991-995 (2022).

\bibitem{Neshev2023} Neshev, D. N. \& Miroshnichenko, A. E. Enabling smart vision with metasurfaces. \textit{Nat. Photon.} \textbf{17}, 26--35 (2023).

\bibitem{Zhu2021} Zhu, S. et al. Integrated photonics on thin-film lithium niobate. \textit{Adv. Opt. Photon.} \textbf{13}, 242--352 (2021).

\bibitem{Jankowski2023} Jankowski, M. et al. Supercontinuum generation by saturated second-order nonlinear interactions. \textit{APL Photonics.} \textbf{8} 116104 (2023).

%31
\bibitem{Guo2023} Guo, Q., et al. Ultrathin quantum light source with van der Waals NbOCl2 crystal. \textit{Nature} \textbf{613} 53--59 (2023).

\bibitem{Wu2017} Wu, L., et al. Giant anisotropic nonlinear optical response in transition metal monopnictide Weyl semimetals. \textit{Nat. Physics} \textbf{13} 350--355 (2017).

\bibitem{Mueller2018} Mueller, T. \& Malic, E. Exciton physics and device application of two-dimensional transition metal dichalcogenide semiconductors. \textit{npj 2D Mater. Appl.} \textbf{2} 29 (2018).

\bibitem{Du2023} Du, L. et al. Moiré photonics and optoelectronics. \textit{Science} \textbf{379} eadg0014 (2023).

\bibitem{Ma2023} Ma, Q. et al. Photocurrent as a multiphysics diagnostic of quantum materials. \textit{Nat. Rev. Phys.} \textbf{5} 170--184 (2023).

%36
\bibitem{Sheffer2023} Sheffer, Y., Queiroz, R. \& Stern, A. Symmetries as the Guiding Principle for Flattening Bands of Dirac Fermions \textit{Phys. Rev. X} \textbf{13} 021012 (2023).

\bibitem{Malard2013} Malard, L. M. et al. Observation of intense second harmonic generation from MoS2 atomic crystals. \textit{Phys. Rev. B.} \textbf{87} 201401 (2013).

\bibitem{Li2013} Li, Y. et al. Probing Symmetry Properties of Few-Layer MoS2 and h-BN by Optical Second-Harmonic Generation. \textit{Nano Lett.} \textbf{13}, 3329–3333 (2013).

\bibitem{Wang2015shg} Wang, G. et al. Giant enhancement of the optical second-harmonic emission of WSe2 monolayers by laser excitation at exciton resonances. \textit{Phys. Rev. Lett.} \textbf{114} 097403 (2015).

\bibitem{Autere2018} Autere, A. et al. Nonlinear Optics with 2D Layered Materials. \textit{Adv. Mater.} \textbf{30}, 1705963 (2018).

\bibitem{Wen2019} Wen, X., Gong, Z. \& Li, D. et al. Nonlinear optics of two-dimensional transition metal dichalcogenides. \textit{Wiley Online Library.} \textbf{1}, 317--337 (2019).

\bibitem{Liu2020} Liu, W. et al. Recent Advances of 2D Materials in Nonlinear Photonics and Fiber Lasers. \textit{Adv. Optical Mater.} \textbf{8}, 1901631 (2020).

\bibitem{Dogadov2022} Dogadov, O. et al. Parametric Nonlinear Optics with Layered Materials and Related Heterostructures. \textit{Laser \& Photonics Reviews.} \textbf{16}, 2100726 (2022).

%44
\bibitem{trovatello2021optical} Trovatello, C. et al. Optical parametric amplification by monolayer transition metal dichalcogenides. \textit{Nat. Photonics.} \textbf{15}, 6--10 (2021).

\bibitem{Zhao2016} Zhao, M. et al. Atomically phase-matched second-harmonic generation in a 2D crystal. \textit{Light: Science \& Applications.} \textbf{5}, e16131--e16131 (2016).

\bibitem{Shi2017} Shi, J. et al. 3R MoS2 with Broken Inversion Symmetry: A Promising Ultrathin Nonlinear Optical Device. \textit{Adv. Mater.} \textbf{29}, 1701486 (2017).

\bibitem{Xu2022} Xu, X. et al. Towards compact phase-matched and waveguided nonlinear optics in atomically layered semiconductors. \textit{Nat. Photon.} \textbf{16} 698--706 (2022).

%48
\bibitem{Weissflog2023ArXiV} Weissflog, M. A. et al. A Tunable Transition Metal Dichalcogenide Entangled Photon-Pair Source. \textit{Nat. Commun.}, \textbf{15}, 7600 (2024).

\bibitem{ArxivTwistPM2023} Hong, H. et al. Twist-phase-matching in two-dimensional materials. \textit{Phys. Rev. Lett.}, \textbf{131}, 233801 (2023).

\bibitem{Shoji97} Shoji, I, Kondo, T., Kitamoto, A., Shirane, M. \& Ito, R. Absolute scale of second-order nonlinear-optical coefficients. \textit{J. Opt. Soc. Am. B}, \textbf{14}, 2268--2294 (1997).

\bibitem{wang2016experimental} Wang, X.- L. et al. Experimental ten-photon entanglement. \textit{Phys. Rev. Lett.} \textbf{117}, 210502 (2016).

\bibitem{MetaRoadmap2024} Kuznetsov, A. I. et al. Roadmap for Optical Metasurfaces. \textit{ACS Photonics} \textbf{11}, 816--865 (2024).
 
\bibitem{Datta2020} Datta, I. et al. Low-loss composite photonic platform based on 2D semiconductor monolayers. \textit{Nat. Photonics.} \textbf{14}, 256--262 (2020).

\bibitem{Bristol2019} Bristol, Paesani, S. et al. Generation and sampling of quantum states of light in a silicon chip. \textit{Nat. Phys.} \textbf{15}, 925--929 (2019).

\bibitem{mannixye2022} Mannix, A. J., Ye, A. et al. Robotic four-dimensional pixel assembly of van der Waals solids. \textit{Nat. Nanotechnol.} \textbf{17}, 361--366 (2022).

\bibitem{Abdelwahab2022} Abdelwahab, I. et al. Giant second-harmonic generation in ferroelectric NbOI$_2$. \textit{Nat. Photonics.} \textbf{16}, 644--650 (2022).

\bibitem{Elshaari2020} Elshaari, A.W., et al. Hybrid integrated quantum photonic circuits. \textit{Nat. Photonics}  \textbf{14}, 285–298 (2020).

\bibitem{silva2017} Li, H. et al. Probing dynamical symmetry breaking using quantum-entangled photons. \textit{Quantum Science and Technology.} \textbf{3}, 015003 (2017).

\end{thebibliography}
\end{document}